# Collective magnetic excitations in mixed-valence $Sm_{0.83}Y_{0.17}S$

P. A. Alekseev,[1] J.-M. Mignot,[2] A. Ochiai,[3] E. V. Nefeodova,[1] I. P. Sadikov,[1] E. S. Clementyev,[1,*]
V. N. Lazukov,[1] M. Braden,[1,4,†] and K. S. Nemkovski,[1]

[1]*Russian Research Centre "Kurchatov Institute", 123182 Moscow, Russia.*

[2]*Laboratoire Léon Brillouin, CEA-CNRS, CEA/Saclay, 91191 Gif sur Yvette, France.*

[3]*Center for Low Temperature Science, Tohoku University, Sendai, Miyagi 980-8578, Japan.*

[4]*Forschungszentrum Karlsruhe, IFP, Postfach 3640, D-76021 Karlsruhe, Germany.*

The magnetic spectral response of black-phase mixed-valence $Sm_{0.83}Y_{0.17}S$ has been measured by inelastic neutron scattering on a single crystal. Two magnetic peaks are observed in the energy range of the $Sm^{2+}$ spin-orbit transition (25–40 meV). Both of them exhibit significant dispersion along the three main symmetry directions, reminiscent of the spin-orbit exciton branch found in pure divalent SmS. The results can be reproduced by a simple phenomenological model accounting for the existence of sizeable Sm-Sm exchange interactions, and a microscopic mechanism is proposed on the basis of the "local-bound-state" theory developed previously for $SmB_6$.

Kondo insulators are unconventional $f$-electron compounds which behave, near room temperature (RT), like an array of independent localized moments interacting with itinerant conduction electrons, while developing clear narrow-gap properties ($E_g \sim 10$ meV) as temperature is reduced.[1, 2] In the latter regime, large many-body renormalization effects are observed. Such behaviors have been reported for materials ($SmB_6$, $Ce_3Bi_4Pt_3$, SmS, $YbB_{12}$, TmSe, etc.) presenting an instability in the occupancy of their $f$ shell, which are hence alternately known as "mixed-valence semiconductors". It has been suggested that the formation of a very small, temperature-dependent gap in the electronic density of states near the Fermi energy may result from the hybridization of the localized $4f$ states with itinerant $5d$-$6s$ states.[3, 4] However, previous inelastic neutron scattering (INS) studies carried out on several compounds of this type ($SmB_6$,[5] $YbB_{12}$[6, 7]) rather point to a local mechanism involving the near neighbors of the $4f$ ion. In $SmB_6$, a gap forms at low temperature in the magnetic excitation spectrum and, at $T = 2$ K, a sharp excitation is observed above the gap at an energy of 14 meV.[5, 8] This extra peak, which exhibits a dramatic decrease in intensity with increasing momentum transfer, was ascribed to a magnetic excitation within an extended "bound state" forming at each Sm site. Such a picture is in line with the theory developed by Mishchenko and Kikoin[9] to explain lattice dynamics anomalies in the mixed-valence (MV) compounds $SmB_6$ and $Sm_{1-x}Y_xS$. The model can actually account for salient features of the magnetic neutron spectra: in particular, the 14-meV excitation corresponds to the reorientation of the total spin and orbital angular momenta associated with an extended component of the MV wave function, expressed as a linear combination of $p$ orbitals from neighboring B atoms.[10] An alternative mechanism for the local-bound-state formation, based on the assumption of a Kondo coupling occurring at $Sm^{3+}$ sites, was proposed by Kasuya.[11]

In the case of $SmB_6$, the new magnetic excitation could be treated as a single-ion property. On the other hand, it might acquire a collective character in other MV systems if the main contribution to the bound state comes from the $d$ orbitals of Sm neighbors. Such a situation could be realized in the MV $Sm_{1-x}Y_xS$ solid solutions. INS experiments performed by Shapiro et al.[12] back in the 1970s revealed that, in pure divalent SmS, the $Sm^{2+}$ singlet-triplet excitation ($J = 0 \rightarrow J = 1$) has a pronounced dispersion ("spin-orbit exciton"), which was successfully modelized, using a mean-field RPA approximation, by taking into account exchange interactions between near-neighbor Sm ions. It was also shown by of Penney and Holtzberg[13] that, upon substitution of trivalent yttrium, extra electrons are transferred into the $5d$ conduction band, and a valence instability of the Sm ions takes place. Therefore, in contrast to pure SmS under pressure, $Sm_{1-x}Y_xS$ alloys become MV already in the "black" ($B$) phase, i.e. before $x$ reaches the critical value for the transition into the collapsed "gold" ($G$) phase.

In this paper, we report on a study of magnetic excitation spectra performed on a $Sm_{0.83}Y_{0.17}S$ ($B$-phase) single crystal, and present evidence for the effect of Sm-Sm interactions on the form of the magnetic response. Experimental information from neutron scattering on the magnetic excitation spectrum of MV Sm in the $Sm_{1-x}Y_xS$ series is still rather scarce, primarily because of prohibitive absorption of thermal neutrons by natural Sm. A $Sm_{0.83}Y_{0.17}S$ single crystal (mass: 1.36 g, volume: $\approx 0.15$ cm$^3$) was thus prepared using isotopically enriched (98.6%) metallic $^{154}$Sm containing less than 0.2% of highly absorbing $^{149}$Sm. Samarium was prereacted with Y and S in a silica tube under vacuum using a conventional electric furnace. The crystal (cubic NaCl structure) was grown by the Bridgmann method in an electron-beam-welded W crucible, then annealed for several days at 1000°C. A spectral analysis confirmed the Y concentration to be $0.17\pm0.03$. The crystal color (dark purple), as well as the room-temperature lattice spacing $a_0 = 5.906(2)$ Å from X-ray diffraction, indicate that the sample is in the $B$ phase.[14] The valence determined from $L_{III}$-edge X-ray absorption spectra for the same material is definitely intermediate ($v = 2.2$), and remains practically constant from room temperature down to 10 K.[15]

Inelastic neutron scattering experiments were carried out at the Laboratoire Léon Brillouin on the 2T1 triple-axis spectrometer (thermal neutron beam), using a double-focusing Cu(111) monochromator and pyrolytic graphite (002) analyzer, with a graphite filter mounted on the scattered beam to suppress higher orders. Energy spectra were measured at different temperatures ($12$ K $\leq T \leq 200$ K) for a fixed final energy of $E_f = 30.5$ meV, yielding a resolution of 2 meV at zero energy transfer. For all three main symmetry directions of the cubic lattice, the magnetic excitation spectra are found to consist of *two* dispersive branches. A double-peak structure is clearly visible in the selection of spectra shown in Fig. 1, measured at the zone center $(2,0,0)$ and for three different $Q$ vectors along [111]. It was checked that no significant difference exists, for equivalent values of $|Q|$ and $|q|$, between spectra measured with longitudinal and transverse scattering configurations. Furthermore, the peak intensities gave no evidence of anisotropy. The dispersions ($a$) and integrated intensities ($b$) of both modes are plotted in Fig. 2. Two important points are to be noted: the dispersions of both modes have an energy minimum at the zone center, and the two modes interchange their intensities as a function of $q$ along all three symmetry directions, the mode with the lower energy being the stronger near zone center and the weaker near zone boundary.

The $Q$-dependence of the *total* integrated intensity of the two modes shows clear magnetic behavior for all three directions studied. The data for $Q \parallel$ [111] are presented in the inset in Fig. 1(a). The experimental intensities have been normalized, for the lowest experimental $Q$-value, to the calculated form factor of the $Sm^{2+}$ intermultiplet transition (no data can be obtained for $Q < 2$ Å$^{-1}$ because of neutron kinematic restrictions). The results have two important features: *i)* the integrated intensity is modulated with the periodicity of the reciprocal lattice, and *ii)* its average decrease seems slightly faster than that expected from the free-ion form factor calculation.[5, 16] The intensity modulation clearly denotes the existence of magnetic correlations between *f* electrons at neighboring lattice sites, in line with the above-mentioned dispersion of the excitation branches. A steeper $Q$-dependence of the intensity, in comparison with a normal ionic form factor, might result from a partial delocalization of the $4f$ electron wavefunctions, as assumed in "local-bound-state" models[10, 11] of mixed valence.

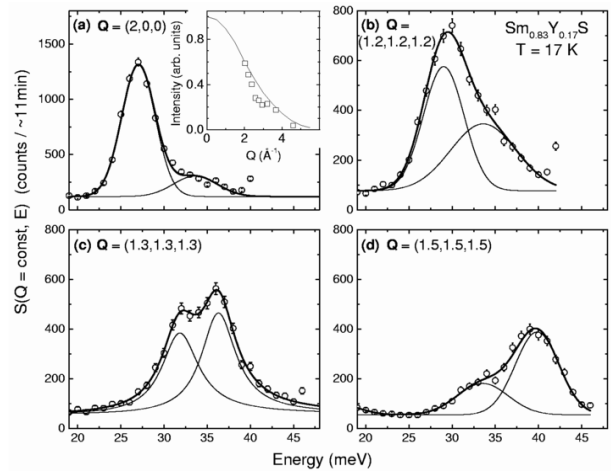

Fig. 1.    Neutron scattering spectra of single-crystal $Sm_{0.83}Y_{0.17}S$ at T = 17 K for different scattering vectors. Lines are fits using two Gaussians [two Lorentzians in (c)]. Inset in (a): Q-dependence of the total integrated intensity of both magnetic peaks for scattering vectors along [111]; intensities were normalized at the lowest experimental Q value to the calculated form-factor dependence (solid line).

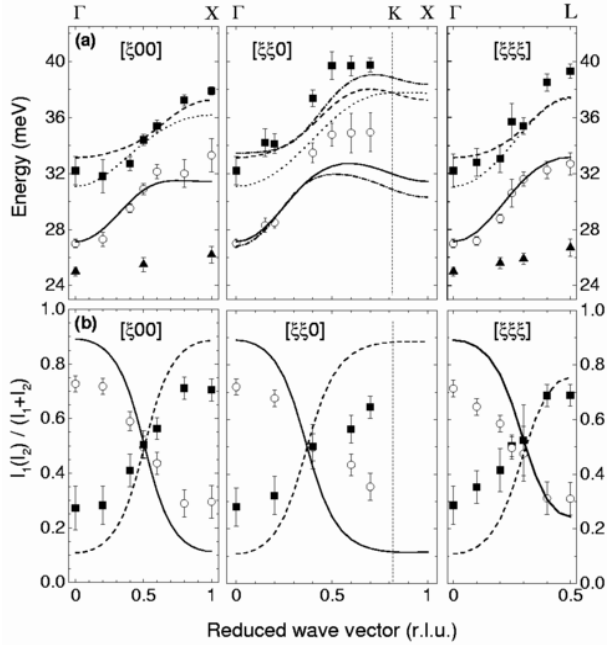

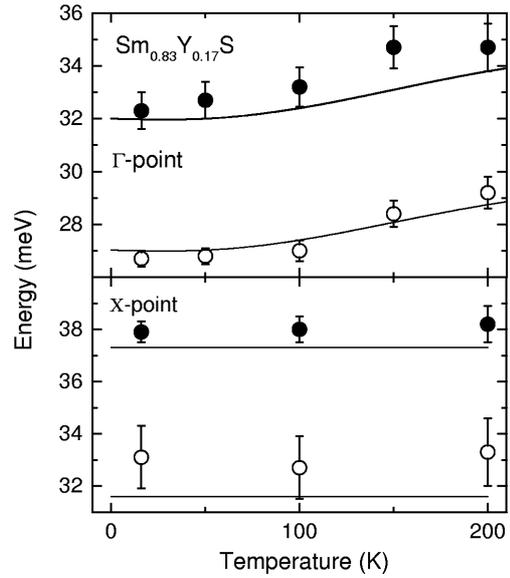

Fig. 2. Energy dispersions (a) and relative intensities (b) along the main symmetry directions of the two magnetic excitation modes in single-crystal $Sm_{0.83}Y_{0.17}S$ at $T = 17$ K. Squares: upper peak; circles: lower peak; triangles: TO phonon branch. Dotted lines in (a): dispersion of the "spin-orbit exciton" in SmS, calculated from the data of Ref. 12. Solid and dashed lines represent the dispersions (a) or relative intensities (b) calculated using the model described in the text for $V_0 = -1.1$ meV and $V_{nn} = 0.35$ meV. The dash-dotted line in (a) for $\boldsymbol{q} \parallel [110]$ is obtained by fitting this particular direction separately.

Fig. 3. Temperature dependence of the energies of the magnetic excitations at the $\Gamma$ point $\boldsymbol{Q} = (2,0,0)$ and the $X$ point $\boldsymbol{Q} = (3,0,0)$; lines represent calculations using the model described in the text .

The temperature dependence of the two magnetic modes was studied between 12 and 200 K. The main effect is a reduction of the overall dispersions with increasing $T$, especially above 100 K. This is evidenced in

Fig. 3, which shows the data for the zone center and the zone boundary in the [100] direction. At the $\Gamma$ point (upper frame) the excitation energies exhibit a significant nonlinear increase with temperature, whereas they remain practically unchanged at the $X$ point (lower frame). The integrated intensities of the peaks were traced as a function of temperature for $\boldsymbol{Q} = (2,0,0)$, and their variations were found to be consistent with that calculated for a simple singlet-triplet energy level scheme.

The comparison of the dispersion curves plotted in Fig. 2 strongly indicates that the upper magnetic branch observed in $Sm_{0.83}Y_{0.17}S$ is analogous to the "spin-orbit exciton" existing in SmS.[12] On the other hand, the existence of a second branch is a unique feature of the alloy system. It is important to note that the temperature dependences of the two branches are very similar, and resemble that of the exciton in the pure compound. This supports the idea that the dispersions are due to the same mechanism, namely indirect Sm-Sm exchange via the $5d$ states. The experimental observation that the two inelastic peaks exchange their intensities when the reduced wave vector varies across the Brillouin zone further suggests a mode coupling effect.

To substantiate these arguments, we introduce a simple phenomenological model based on the assumption of two magnetic modes hybridizing with each other, whose dispersions can be represented in the formalism of the singlet-triplet MF-RPA model. In the context of the "local-bound-state" picture of Ref. 10, these modes might be assigned, respectively, to the "parent" $Sm^{2+}$ ($J = 0 \rightarrow J = 1$) spin-orbit excitation, and to the renormalized one ($J^* = 0 \rightarrow J^* = 1$) associated with the quantum-mechanical MV ground state.

Following Ref. 12, the dispersions of the two magnetic modes ($\lambda = 1,2$ for the higher- and lower-energy excitation, respectively) in the non-interacting limit is given by

$$\hbar\omega_\lambda(\mathbf{q}) = \Delta_\lambda\left[1 - 4\left|\left\langle{}^7F_1\left|\mathbf{S}\right|{}^7F_0\right\rangle\right|^2 R(T)\frac{J(\mathbf{q})}{\Delta_\lambda}\right]^{1/2} \quad (1)$$

where $J(\mathbf{q})$ is the Fourier transform of the Sm-Sm indirect exchange interaction and $R(T)$ a temperature factor. In calculating $J(\mathbf{q})$, exchange constants up to the third nearest neighbors are considered, and their values are assumed to be the same for both modes and equal to those determined for pure SmS ($J_1 = 0.043$ meV, $J_2 = 0.025$ meV, $J_3 = -0.003$ meV).[12] We take $\Delta_1 = 36.2$ meV, corresponding to the single-ion energy of the $Sm^{2+}$ $J = 0 \rightarrow J = 1$ transition, and a starting value of $\Delta_2 = 32.5$ meV from the empirical relation between the "local bound state" energy and the Sm valence in $Sm_{1-x}La_xB_6$ alloys.[17]

The mixing interaction between the two modes is written as the sum of an on-site (dominant) term $V_0$ and a smaller contribution $V_1(\mathbf{q}) = \sum_j V_{nn} \exp\left[i\mathbf{q}(\mathbf{r}_j - \mathbf{r}_i)\right]$ due to the hybridization ($V_{nn}$) of one mode from a given Sm with the other mode from a neighboring site belonging to the first coordination sphere.

The above interaction can be treated in standard second-order perturbation theory, yielding the energies and intensities for the hybridized modes. The energies were fitted to the experimental dispersion curves separately for each symmetry direction using only two adjustable parameters $V_0$ and $V_{nn}$ (dash-dotted lines in Fig. 2($a$) for the [110] direction). The resulting values for $V_0$ and $V_{nn}$ were found to be consistent within 20%, so that all three directions could be fitted simultaneously for $V_0 = -1.1$ meV and $V_{nn} = 0.35$ meV (solid and dashed lines in Fig. 2($a$)). Further varying $\Delta_2$ from its initial value of 32.5 meV did not improve the agreement significantly. The relative intensities of the two modes, calculated for the same values of the parameters (assuming the single-ion values of the magnetic dipole matrix elements for both excitations to be equal) are compared with the experimental data in Fig. 2($b$). The model correctly predicts that the lower mode dominates at the zone center, and reproduces the crossing of the curves about halfway to the zone boundary. This behavior basically reflects the fact that the change in the energies of the modes due to hybridization is a function of $|V(\mathbf{q})|^2$ whereas their relative intensities depend on the value and sign of $V(\mathbf{q})=V_0+V_1(\mathbf{q})$. The latter expression goes through zero at an intermediate value of $q$, for which the negative on-site and positive inter-site contributions cancel out. The calculated temperature dependence of the excitation energies also agrees quite well with the experimental data plotted in Fig. 3.

Our assumption that the exchange constants in $Sm_{0.75}Y_{0.25}S$ remain approximately the same as in pure SmS is at variance with the interpretation of previous bulk susceptibility measurements, in which an enhancement of Sm-Sm exchange due to the appearance of interactions via conduction electrons was proposed[18, 19] to be responsible for the increase observed in the magnetic Van-Vleck susceptibility of $Sm_{1-x}La_xS$ and $Sm_{1-x}Y_xS$ alloys with increasing $x$ ($x \leq 0.23$). However, the enhancement of the bulk susceptibility may have a different origin, namely the appearance of an extra magnetic mode, as observed in our measurements, with low energy and a strong intensity at the $\Gamma$ point.

As shown above, a simple phenomenological model can account reasonably well for the interaction between the two branches of excitations observed experimentally. On the other hand, it cannot answer the question of the origin of an extra mode as compared to pure SmS. In a previous time-of-flight INS study on polycrystalline $Sm_{0.75}Y_{0.25}S$,[20] the spectral response with a complex fine structure measured for energies between 25 and 40 meV was tentatively ascribed to $Q$-space averaging of two dispersive excitations, which were assumed to result from the resonant coupling of a spin-orbit exciton with a TO phonon of the same symmetry occurring in the same energy range. However, the temperature renormalization of the phonon energy was not taken into account. In our measurements, on the other hand, the role of phonons does not seem to be significant in view of the $Q$-dependence presented in Fig. 1. To clarify this point, we have searched for optic phonon modes in our crystal over the same energy range where the magnetic double-peak structure is observed. The intensity of the TO branch is rather weak but it could nonetheless be measured along the [100] and [111] directions for $T = 12$ and 300 K. Within experimental accuracy, the energies of the modes are the same at both temperatures. The dispersion curves are shown in Fig. 2(a) (triangles) together with those of the magnetic excitations. As can be seen, the phonon branches lie at much lower energies (25–27 meV) than the magnetic contribution. Therefore, at least for the present ($B$-phase) composition, we consider unlikely that the dispersive modes originate from strong magnetoelastic coupling.

On the other hand, an interpretation based on a "local-bound-state" model of the MV phenomenon provides a rather natural explanation for the existence of the lower branch. As noted in the introduction, if the extended component of the MV wave function consists principally of $d$ orbitals from the Sm neighbors, the renormalized spin-orbit excitation corresponding to the 14-meV peak in $SmB_6$ may become dispersive by the same mechanism as the usual spin-orbit transition. Furthermore, as shown in Ref. 10, the extra mode has the same symmetry as the conventional $J = 0 \rightarrow J = 1$ spin-orbit transition, which provides additional support to the mode-mixing effect introduced phenomenologically in the model. The emerging picture could be that of a collective state in which the loosely bound electron is distributed among several Sm sites, in contrast to the local state obtained when the $f$-electron density remains located on the first coordination shell of the rare-earth site.

In summary, two dispersive excitations have been observed in the magnetic response of the mixed-valence ($B$-phase) alloy $Sm_{0.83}Y_{0.17}S$. A simple phenomenological model based on two interacting magnetic modes can explain consistently the dispersions and relative intensities of these excitations for the three high-symmetry directions. Whereas the upper branch is clearly related to the "spin-orbit exciton" previously reported for pure SmS, the second one could arise from a "collective bound state", similar to that proposed previously for MV $SmB_6$, but with significant cooperative character. The present results provide clear evidence that Sm-Sm exchange interactions play a major role in the formation of the MV state. Further experiments on the $G$-phase $Sm_{1-x}Y_xS$ alloys should be undertaken to study the evolution of the magnetic response as a significant fraction of the Sm $4f$ electrons is promoted into extended $5d$ states.

We wish to thank A.S. Mishchenko for fruitful discussions. The work was supported by the RFBR grant 99-02-16522 and the Russian State Program NICS. P.A.A. is grateful to the LLB for hospitality and financial support.

---

*Present address: Technische Universität München, Physik E21, James Franck Str. 1, D-85748 Garching, Germany.

†Present address: II. Physikalisches Institut, Universität zu Köln, Zülpicher Straße 77, D-50937, Köln, Germany.